\documentclass[aps,prl,preprint,groupedaddress,showpacs]{revtex4}

\usepackage{graphicx}
\usepackage{amsmath}
\usepackage{epsfig}

\begin{document}


\title{Laser-driven shock acceleration of monoenergetic ion beams}

\author{F. Fiuza}
\email[Electronic address: frederico.fiuza@ist.utl.pt]{}

\author{A. Stockem}
\author{E. Boella}
\author{R. A. Fonseca}

\author{L. O. Silva}
\email[Electronic address: luis.silva@ist.utl.pt]{}
\affiliation{GoLP/Instituto de Plasmas e Fus\~ao Nuclear - Laborat\'orio Associado, Instituto Superior T\'ecnico, Lisboa, Portugal}

\author{D. Haberberger}
\author{S. Tochitsky}
\author{C. Gong}
\author{W. B. Mori}
\author{C. Joshi}
\affiliation{Department of Electrical Engineering, University of California, Los Angeles, California 90095, USA}
\affiliation{}

\date{\today}

\begin{abstract}
We show that monoenergetic ion beams can be accelerated by moderate Mach number collisionless, electrostatic shocks propagating in a long scale-length exponentially decaying plasma profile. Strong plasma heating and density steepening produced by an intense laser pulse near the critical density can launch such shocks that propagate in the extended plasma at high velocities. The generation of a monoenergetic ion beam is possible due to the small and constant sheath electric field associated with the slowly decreasing density profile. The conditions for the acceleration of high-quality, energetic ion beams are identified through theory and multidimensional particle-in-cell simulations. The scaling of the ion energy with laser intensity shows that it is possible to generate $\sim 200$ MeV proton beams with state-of-the-art 100 TW class laser systems.
\end{abstract}



\maketitle

The electric and magnetic fields excited in a plasma by laser pulses allow for the acceleration of ions to high energies over short distances \cite{bib:malka}. Such accelerated beams are of interest for a broad range of potential applications: cancer therapy \cite{bib:bulanov,bib:ledingham}, isotope generation \cite{bib:spencer}, proton radiography \cite{bib:borghesi}, and fast ignition \cite{bib:roth}. Many applications require low energy spread (1-10\% FWHM) and low emittance ion beams. Radiotherapy applications require a relatively high beam energy, in the range of $100-300$ MeV/a.m.u. \cite{bib:linz}. 

A significant effort has been devoted in recent years to determining the optimal parameters for ion acceleration. The two most studied mechanisms are target normal sheath acceleration (TNSA) \cite{bib:tnsa} and radiation pressure acceleration (RPA) \cite{bib:rpa}. However, the production of high-energy and high-quality ions beams remains a challenge. A third mechanism for accelerating ions to high energies in a laser-produced plasma is shockwave acceleration (SWA) \cite{bib:denavit,bib:silva,bib:d'humieres,bib:sorasio,bib:macchi}. Here, a shock propagates in the plasma with a velocity $v_{sh}$ which can reflect ions from the background plasma to a velocity $v_{ions} \sim 2 v_{sh}$. Previous theoretical and numerical studies focused on the conditions under which shocks are formed in solid targets using extremely high laser intensities and the accelerated ion spectrum was broad \cite{bib:silva,bib:d'humieres}. Motivated by recent experimental results on monoenergetic acceleration of protons \cite{bib:haberberger}, we consider SWA in near critical plasma density targets at modest laser intensities.

We first derive the conditions for ion reflection from an electrostatic collisionless shock as a function of the initial density and temperature profile, and for relativistic temperatures. We characterize and optimize a new acceleration regime which combines SWA with a special regime of TNSA. In this regime, the shock propagates through an extended exponentially decreasing density profile, where the sheath fields are small and constant \cite{bib:mora}, and the shock-accelerated ions maintain a narrow energy spread. We show, through multi-dimensional particle-in-cell (PIC) simulations, that shocks satisfying these conditions can be driven in laser-plasma interactions at near critical density and confirm our theoretical scaling of the final proton energy with laser intensity (or normalized vector potential $a_0$). Our results establish the conditions to achieve the beam energy and quality required for medical applications in near future experiments with readily available laser systems ($a_0 \sim 10$).

\begin{figure}[h]
\begin{center}
\includegraphics[width=0.9\textwidth]{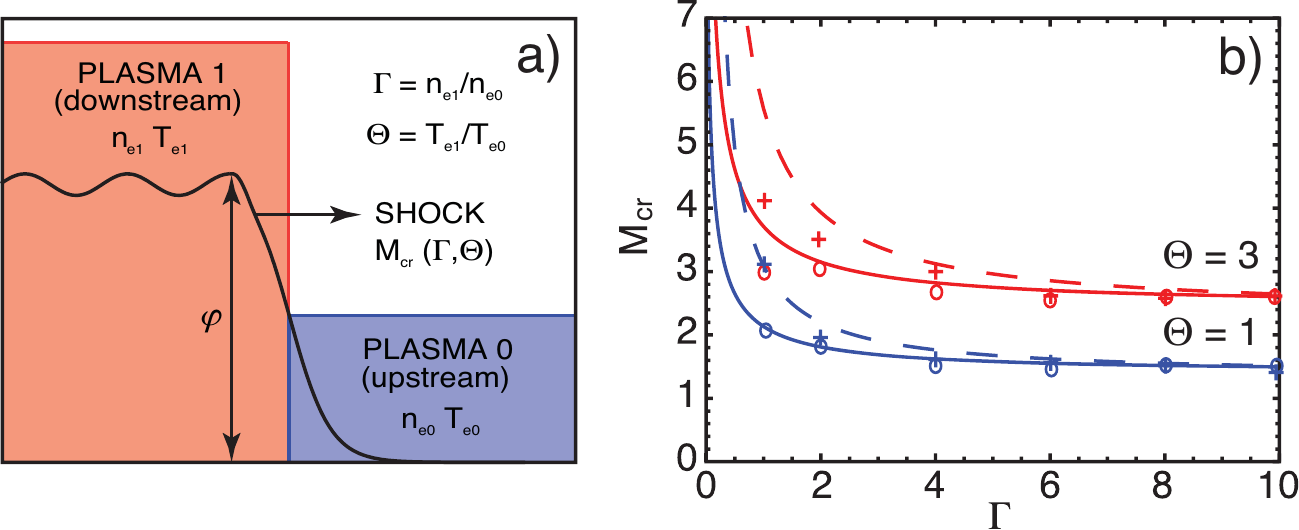}
\caption{\label{fig:theory} a) Schematic representation of the interaction between two plasma slabs with different temperature and density (red and blue), which leads to shock formation (black line represents the electrostatic potential) and ion reflection. b) Critical Mach number for ion reflection as a function of the density ratio $\Gamma$ and temperature ratio $\Theta$ between the two plasma slabs/regions, for $T_{e\,0} = 1$ keV (dashed line \cite{bib:sorasio}) and $T_{e\,0} = 1.5$ MeV (solid line Eq. \eqref{eq:mcrit}). The symbols indicate the simulation values for the nonrelativistic (+) and relativistic (o) electron temperatures, obtained by measuring the speed of the shock structure (density jump or electrostatic field) when ion reflection is observed.}
\end{center}
\end{figure}

To study shock formation and ion acceleration, we consider the interaction of two plasma slabs (denoted by plasma 1 and plasma 0) with electron temperature ratio $\Theta = T_{e1}/T_{e0}$ and density ratio $\Gamma = n_{e1}/n_{e0}$ (Fig. \ref{fig:theory} a). Electrostatic shock structures can be generated as a result of the expansion of plasma 1 (downstream - region behind the shock) into plasma 0 (upstream - region ahead of the shock) \cite{bib:sorasio, bib:shamel}, with the dissipation provided by the trapped particles behind the shock and, for strong shocks, by the ion reflection from the shock front \cite{bib:sagdeev}. The nonrelativistic theory, whereby an electrostatic shock is supported by regions/slabs of arbitrary temperature and density ratios, has been outlined in \cite{bib:sorasio}. We have generalized this theoretical framework for relativistic electron temperatures \cite{bib:stockem} and use it here in order to study the optimal conditions for ion reflection by the shock. The generalized nonlinear Sagdeev potential is given by $\Psi (\varphi) = P_i (\varphi, M) - P_{e\,1} (\varphi, \Theta, \Gamma, \mu_{e\,0}) - P_{e\,0} (\varphi, \Gamma, \mu_{e\,0})$, where $P_{e\,1} (\varphi, \Theta, \Gamma, \mu_{e\,0}) = \Gamma \Theta/(1+\Gamma) \Big\{(\mu_{e\,0}/\Theta)/K_1 [\mu_{e\,0}/\Theta] \Big[\int_1^\infty d\gamma e^{-\mu_{e\,0} \gamma/\Theta} \sqrt{(\gamma+\varphi/\mu_{e\,0})^2-1} + e^{-\mu_{e\,0}/\Theta} (\sigma \sqrt{\sigma^2-1}  - \mathrm{Log}[\sigma + \sqrt{\sigma^2-1}]) \Big]-1 \Big\}$ is the downstream electron pressure, $P_{e\,0} (\varphi, \Gamma, \mu_{e\,0}) = 1/(1+\Gamma) \left\{ \left [(\mu_{e\,0}/K_1 [\mu_{e\,0}])\int_1^\infty d\gamma e^{-\mu_{e\,0} \gamma} \sqrt{(\gamma+\varphi/\mu_{e\,0})^2-1}\right] -1\right\}$ is the upstream electron pressure, and $P_i (\varphi, M) = M^2(1-\sqrt{1-2\varphi/M^2})$ is the ion pressure for the assumption of cold ions and relativisitic Maxwellian electrons, with $\mu_{e\, 0} = m_e c^2/k_B T_{e\, 0}$ the inverse of the normalized electron temperature. Here, $M = v_{sh}/c_{s\,0}$ is the shock Mach number, $c_{s\,0} = (k_B T_{e0}/m_i)^{1/2}$ is the upstream sound speed, $\varphi = e\phi/k_B T_{e\,0}$ is the electrostatic potential energy across the shock front normalized to the upstream thermal energy, $k_B$ is the Boltzmann constant, $m_i$ and $m_e$ are the ion and electron mass, $\sigma = 1+\varphi/\mu_{e\,0}$, and $K_1$ is the modified Bessel function of the second kind. In the ultra-relativistic limit, $\mu_{e\, 0} \ll 1$, $P_{e\,1} (\varphi, \Theta, \Gamma, \mu_{e\,0}) = \varphi \Gamma[\mu_{e\,0}(1-\varphi/\Theta) + \varphi + \Theta]/[(1+\Gamma)\Theta]$, and $P_{e\,0} (\varphi, \Theta, \Gamma, \mu_{e\,0}) = \varphi (1-\mu_{e\,0})/(1+\Gamma)$. Shock solutions can be found for $\Psi (\varphi) < 0$ \cite{bib:tidman}. Ion reflection from the shock front will occur when the electrostatic potential across the shock exceeds the kinetic energy of the upstream ions, {\it i.e.} $\varphi_{cr} = M_{cr}^2/2$, which defines the critical Mach number
\begin{equation} 
M_{cr} = \sqrt{2 \Theta \left( \frac{1+ \mu_{e\,0}}{\Gamma (1- \mu_{e\,0}/\Theta )} +1\right)}.
\label{eq:mcrit}
\end{equation}

Two-dimensional (2D) PIC simulations, performed with OSIRIS 2.0 \cite{bib:fonseca}, of the interaction of two semi-infinite plasma slabs with different temperature and density ratios are in good agreement with theory (Fig. \ref{fig:theory} b). Ion reflection can occur for moderate Mach number shocks provided that $\Gamma \gg 1$ and $\Theta \sim 1$. At high density ratios $\Gamma \ge 4$, the expansion of the two slabs (initially at rest) is sufficient to form the shock and reflect the ions. At lower density ratios, the plasma slabs need to have an initial relative drift in order to reach $M_{cr}$ for ion reflection.

\begin{figure}[h]
\begin{center}
\includegraphics[width=0.9\textwidth]{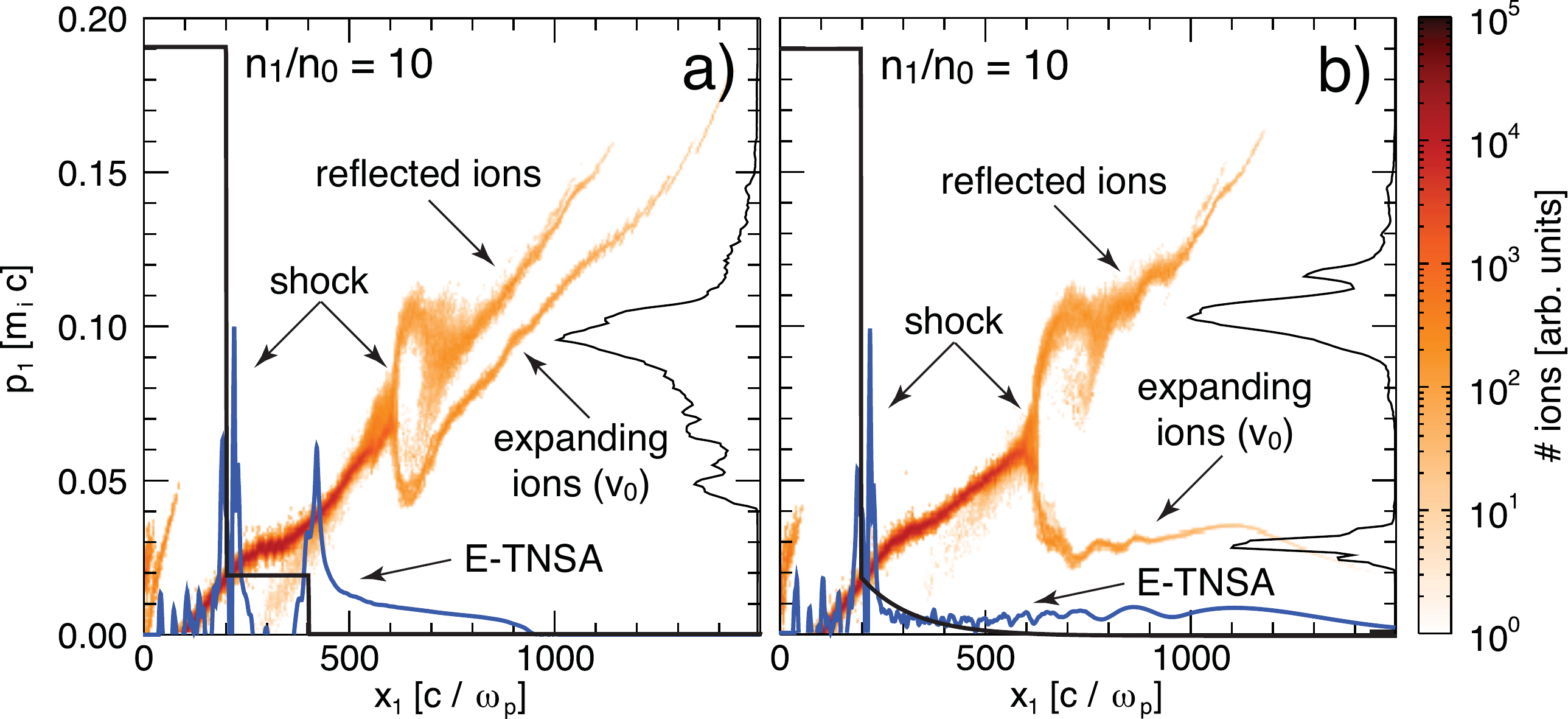}
\caption{\label{fig:slabs} Ion phase-space at $7700 ~\omega_{p1}^{-1}$ after the interaction of two plasma slabs with initial temperature of 1.5 MeV, $\Theta = 1$, and $\Gamma = 10$. In a) a flat density profile is used for the low density slab, whereas in b) it is replaced by an exponential profile. The black lines indicate the initial plasma density profile and the blue lines indicate the early ($t = 560 ~\omega_{p1}^{-1}$) longitudinal electric field. The thin black lines indicate the integrated ion spectrum ahead of the shock.}
\end{center}
\end{figure}

In more realistic configurations, where finite slabs are considered, it is important to address the role of competing accelerating fields. As hot electrons expand into vacuum, TNSA fields will develop at the plasma-vacuum interface accelerating the upstream ions to a given velocity $v_0$. The shock will then reflect the upstream ions to a velocity $v_{ions} \simeq 2 M_{cr} c_{s\,0} + v_0$. This is shown in Fig. \ref{fig:slabs} by simulating the interaction of two finite plasma slabs with $\Gamma = 10$ and $\Theta=1$. For an abrupt plasma-vacuum transition, the electrostatic field in the sheath at the rear side of the upstream plasma introduces a chirp in $v_0$ \cite{bib:silva}, broadening the ion energy spectrum as typical of TNSA \cite{bib:tnsa, bib:mora} (Fig. \ref{fig:slabs} a). This sheath field can be controlled by using an exponential plasma profile with scale length $L_g$, which is characterized by a constant electric field at early times ($t \ll 4 L_g/c_{s\,0}$) \cite{bib:grismayer} given by
\begin{equation} 
E_{TNSA} = \frac{k_b T_{e\,0}}{e L_g},
\label{eq:efield}
\end{equation}
as illustrated in Fig. \ref{fig:slabs} b, where we replace the low density slab of Fig. \ref{fig:slabs} a with an exponentially decreasing profile. It can be observed that the shock accelerated ions are able to cross the sheath region while preserving their narrow energy spread, thus indicating a configuration suitable for the generation of monoenergetic ion beams.

\begin{figure}[h]
\begin{center}
\includegraphics[width=0.9\textwidth]{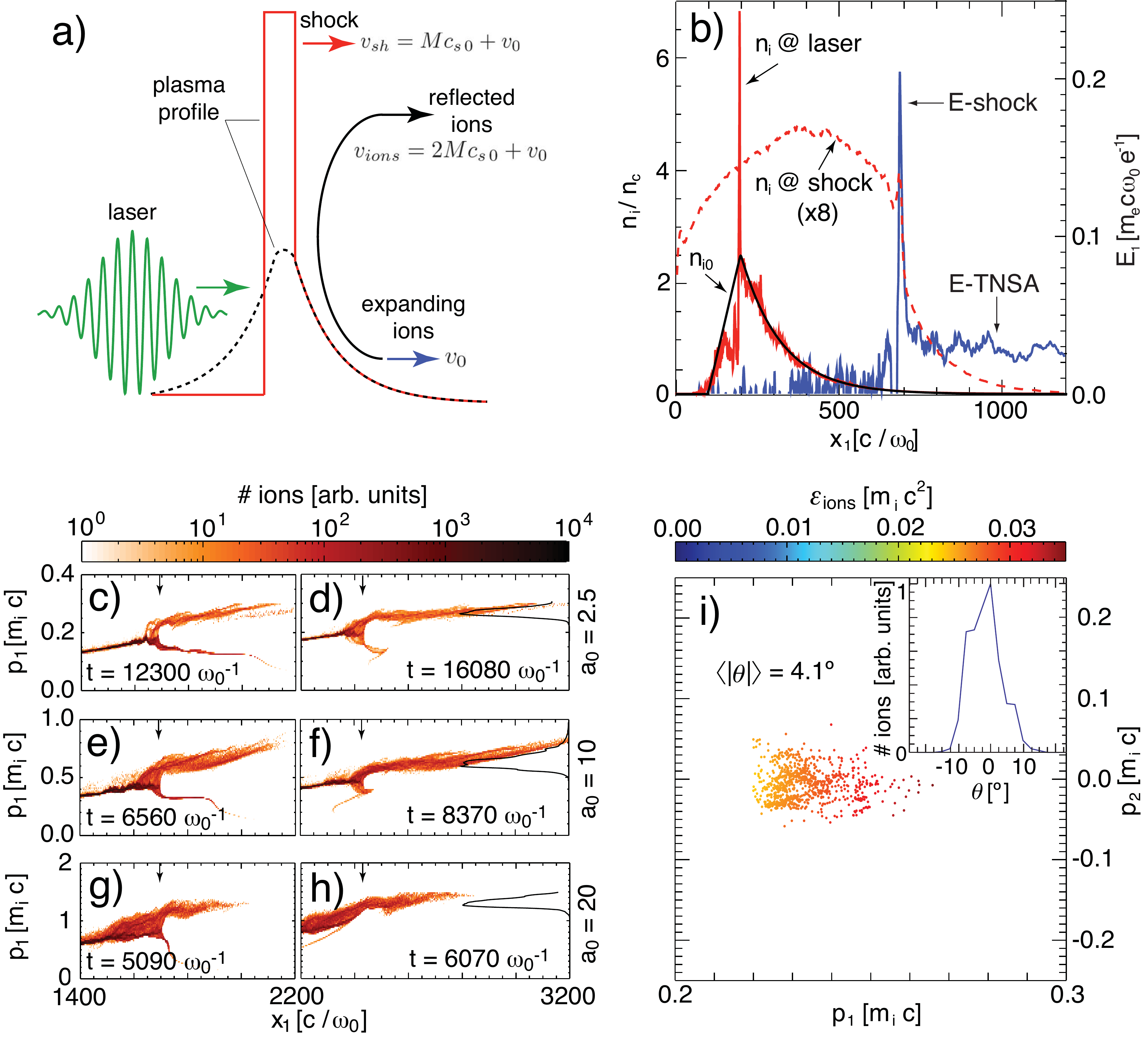}
\caption{\label{fig:shock} a) Schematic representation of SWA driven by the interaction of a laser (green) with a near critical density plasma (dashed). The laser heats up the plasma electrons and steepens the density profile at the critical density (red), driving a shock which reflects the upstream expanding ions. b) Transversely averaged density and field structure of the shock driven by a laser with $a_0 = 2.5$. The density profile is shown at $t  = 0$ (black), at the interaction of the peak of the laser with the critical density (red), and after shock formation (dashed red), $t = 6560 ~\omega_{0}^{-1}$, together with the longitudinal electric field (blue). The uniform TNSA field is in good agreement with Eq. \eqref{eq:efield} for the measured $T_{e0} = 1.6$ MeV ($E_{TNSA} = 0.025 m_e c \omega_0 / e$). c - h) Ion phase-space evolution for $a_0 = 2.5$ (c, d), 10 (e, f), and 20 (g, h). Shock position is indicated by the arrows and the black lines indicate the final integrated spectrum of the reflected ions. d) Momentum distribution of the accelerated ion beam for a finite laser spot size $W_0 = 16 \lambda_0$.}
\end{center}
\end{figure}

The conditions for shock formation and monoenergetic ion acceleration can be obtained in practice from the interaction of a moderate intensity laser pulse with a tailored plasma density profile (see Fig. \ref{fig:shock} a). For near critical density plasmas a significant fraction ($> 20$\% \cite{bib:wilks,bib:tsung}) of the laser energy can be absorbed maximizing electron heating, and therefore ion acceleration. In the relativistic regime, the electron temperature, $3 k_B T_e = \epsilon_e$, can be estimated by equating the plasma electron energy density to the absorbed laser energy density, $3 a_0 n_c L_{target} k_B T_e = \eta I \tau_{laser}$, where $\eta$ is the absorption efficiency and the relativistically corrected critical density $a_0 n_c$ has been used, yielding
\begin{equation} 
T_e [\mathrm{MeV}] \simeq 0.026 \eta a_0 \frac{\tau_{laser} [\mathrm{ps}]}{L_{target} [\mathrm{mm}]}.
\label{eq:temperature}
\end{equation}
For typical picosecond scale laser pulses with relativistic intensities, $a_0 > 1$, and target size $L_{target} < 1$ mm, strong heating to MeV temperatures can occur, leading to high shock velocities and high reflected ion energies. 

For the production of a monoenergetic ion beam, the shock velocity should be uniform, which relies on uniform heating of the plasma electrons. That is achieved by allowing the heated electrons to recirculate in the target before shock formation \cite{bib:silva, bib:mackinnon}. From \cite{bib:forslund} the shock formation time is $\sim 4 \pi/\omega_{pi}$ for $M \sim 1$. For MeV electrons to recirculate at least once in the target, the target size should be limited to $L_{target} < (m_i/m_e)^{1/2} \lambda_0$ for critical density targets. The shock will efficiently reflect a uniform ion population if the expanding ion velocity, $v_0 = (c_{s\,0}^2/L_g)t$, is much smaller than the shock velocity by the time the shock is formed, {\it i.e.} if $L_g \gg 2 (c_{s\,0}^2/{v_{sh}} c) (m_i/m_e)^{1/2} \lambda_0$. For a symmetric target expansion ($L_{target} \leq 2 L_g$) and low Mach number shocks ($M \gtrsim 1$), the optimal target scale length for uniform electron heating and ion reflection is
\begin{equation} 
L_{g\,0} \approx \frac{\lambda_0}{2} \left(\frac{m_i}{m_e}\right)^{1/2}.
\label{eq:optimal}
\end{equation}
Stable SWA requires a shock width (which is close to the laser spot size $W_0$) larger than the transverse expansion of the plasma, at $c_s$, during the acceleration. Assuming an isothermal expansion, this condition yields $W_0 \gtrsim L_{g\,0}/M_{cr}$.

The final ion energy is given by the combination of SWA with the uniform expansion of the upstream plasma. The final relativistic ion velocity is $v_{ions} = (v_{sh}' + v_{0})/(1+v_{sh}' v_{0}/c^2)$, where $v_{sh}' = (2 M c_{s \,0})/(1+ M^2 c_{s \,0}^2/c^2)$ is the velocity of the reflected ions in the upstream frame and $v_{0}$ is the upstream velocity at the shock acceleration time $t_{acc}$. Taylor expanding $v_{ions}$ for $c_{s \,0}/c \ll 1$, the proton energy for optimal conditions is
\begin{equation} 
\epsilon_{ions} [\mathrm{MeV}] \simeq 2 M_{cr}^2 T_{e \,0} [\mathrm{MeV}] + M_{cr}\frac{c t_{acc}}{L_{g\,0}}\frac{(2 T_{e \,0} [\mathrm{MeV}])^{3/2}}{(m_i/m_e)^{1/2}} + \left[\left(\frac{c t_{acc}}{L_{g\,0}}\right)^2 + 4 M_{cr}^4\right]\frac{(T_{e \,0} [\mathrm{MeV}])^{2}}{m_i/m_e}.
\label{eq:scaling}
\end{equation}

To explore the proposed generation of high-quality ion beams to 10s $-$ 100s MeV from a laser-driven electrostatic shock we have performed 2D OSIRIS simulations. The simulation box size is $3840 \times 240 ~(c/\omega_0)^2$ with $12288 \times 768$ cells, $9 - 36$ particles per cell per species, cubic particle shapes, and current smoothing. We start by modeling the interaction of a Gaussian laser pulse, duration of $1885 \omega_0^{-1}$ (FWHM) and infinite spot size with a pre-formed electron-proton plasma profile with a linear rise over $10 \lambda_0$, and exponential fall with $L_g = 20 \lambda_0$ (according to Eq. \eqref{eq:optimal}). Increasing laser intensities ($a_0 = 2.5 - 20$) have been used and the peak density of the plasma ($n_p/n_c = 2.5-10$) changed to compensate for an increased relativistic transparency. The laser pulse is highly absorbed as it interacts with the near critical density plasma ($\sim 60$ \% absorption) and stopped at the critical density surface causing a local steepening and leading to a density spike with $3 - 4$ times the background density (Fig. \ref{fig:shock} b). This density spike leads to the onset of shock formation, around $t \sim 4500 ~\omega_0^{-1}$ ($530 ~\omega_0^{-1}$ after the laser interaction finished). The shock structure has a strong localized electric field at the shock front, with a measured thickness of $L_{sh}  \sim 4 \lambda_D = 10 c/\omega_0$, where $\lambda_D = \sqrt{k_B T_e/4 \pi n_p e^2}$ is the Debye length, much smaller than the mean free path for particle collisions ($L_{sh} \ll \lambda_{e\,i} \sim c/\nu_{e\,i} \sim 2 \times 10^8 \lambda_D$, $\lambda_{i\,i} \sim c_{s\,0}/\nu_{i\,i} \sim 2\times 10^2 \lambda_D$, for $T_e = 1$ MeV, $T_i = 100$ eV, and $n_e = n_i = 10^{21}$ cm$^{-3}$).

After the formation, the shock maintains a near uniform velocity, with a Mach number in the upstream reference frame $M = (v_{sh} - v_0)/c_{s\,0} \sim 1.7$, in good agreement with the theoretical $M_{cr}$ for large $\Gamma$ and $\Theta \sim 1$, $M_{cr} \sim 1.5 - 1.8$ (Fig. \ref{fig:theory} b). The temporal ion phase-space evolution is shown in Figures \ref{fig:shock} c-h for the cases of $a_0 = 2.5$, 10, and 20, where it is possible to observe the self-similarity of the interaction. The shock is able to reflect the cold, uniformly expanding ions from the back of the target generating a beam with an energy of 31, 165, and 512 MeV, respectively. The upstream ion temperature measured during the acceleration is relatively small (100 keV for $a_0 = 2.5$ and 1 MeV for $a_0 = 20$). The uniform shock and upstream velocities lead to a reflected ion beam with a low total energy spread of $\sim 10 \%$ (FWHM) and an average slice energy spread of 4 \% (FWHM). The laser to ion beam energy conversion efficiency is measured to be $2-3$ \% in all simulations. The fraction of upstream ions reflected by the shock ranges between 10-20\%. Assuming cylindrical symmetry, the total number of accelerated ions as inferred from the simulation is given by $N_{ions} \sim 10^{10} (W_0 [\mu \mathrm{m}])^2/\lambda_0 [\mu \mathrm{m}]$, where $W_0$ is the laser spot size, ideal for most applications. For instance, in radiotherapy $\sim 10^8$ ions per bunch are used in multi-shot treatment and $\sim 10^{11}$ ions per bunch in single shot treatment \cite{bib:bulanov,bib:linz}.

We have confirmed that our picture for SWA is still valid for a finite laser spot size by performing 2D simulations under the same conditions, with $a_0 = 2.5$ and a super-Gaussian spot size $W_0 = 16 \lambda_0$. A monoenergetic ion beam with 28 MeV and a narrow energy spread of 9 \% similar to that shown in Fig. \ref{fig:shock} d was produced. This ion beam has a small divergence of 4$^\circ$ half angle (Fig. \ref{fig:shock} i).

It is important to note that the acceleration of ions by the shock occurs after the laser has fully interacted with the plasma; the accelerated beam properties do not depend on the exact laser pulse profile and are not significantly affected by laser-plasma instabilities in the front of the target, such as filamentation. The beam quality depends mainly on the target profile. Simulations performed outside the optimal parameter range (Eq. \eqref{eq:optimal}) led to an ion bunch with a larger energy spread and/or less energy (using $L_g = 40 \lambda_0$ resulted in a 17 MeV ion bunch with an energy spread of $\sim 30$ \%).

\begin{figure}
\begin{center}
\includegraphics[width=0.9\textwidth]{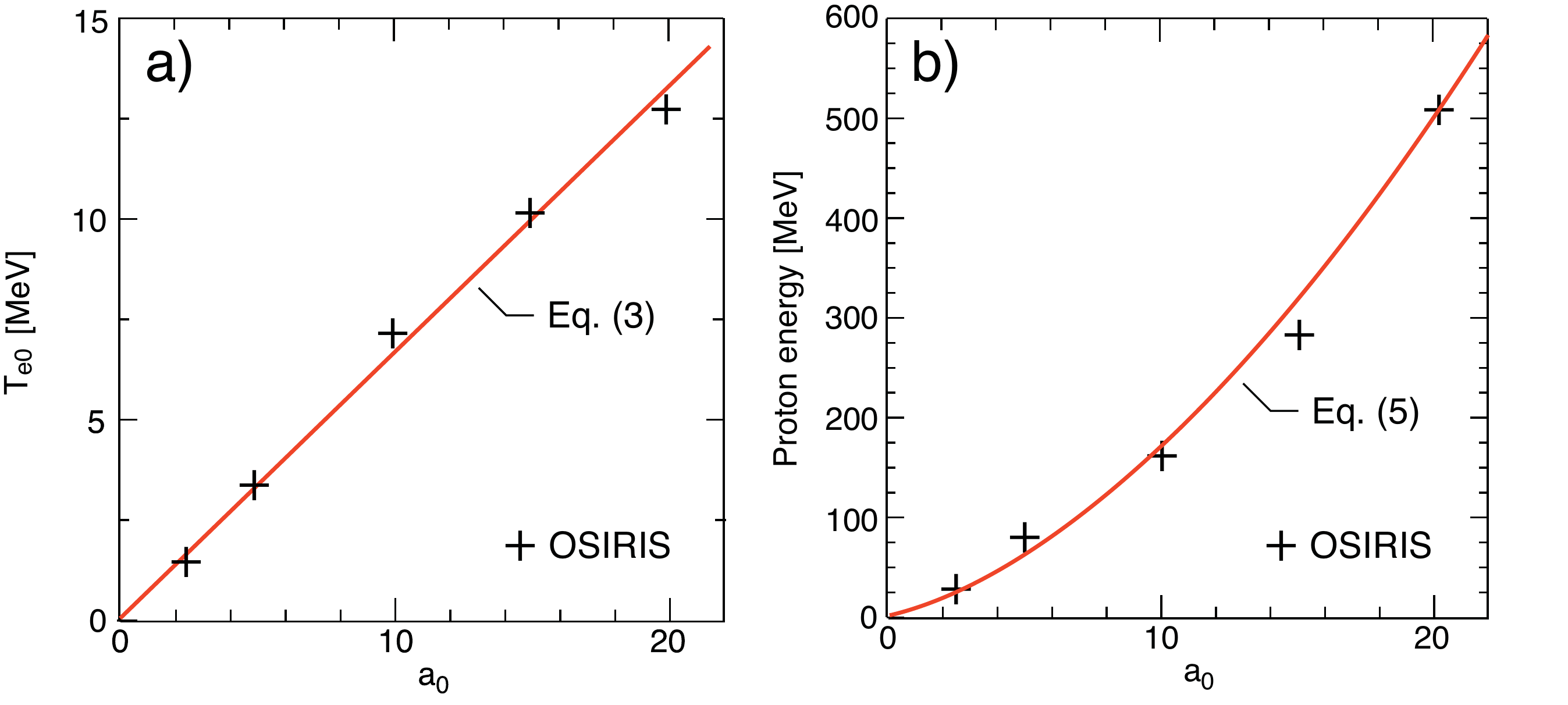}
\caption{\label{fig:scaling} Scaling of a) electron temperature and b) shock accelerated proton energy with laser $a_0$. The electron energy distribution obtained in OSIRIS is well fitted by a relativistic Maxwellian. The obtained scalings are consistent with Eq. \eqref{eq:temperature} for $\eta = 0.51$ and Eq. \eqref{eq:scaling} for $t_{acc} = 5500 \omega_0^{-1}$, respectively.}
\end{center}
\end{figure}

The scaling of this scheme with the laser/plasma parameters was investigated by comparing the simulation results with our theoretical estimates. The electron temperature is observed to scale linearly with the laser amplitude (Fig. \ref{fig:scaling} a), which is consistent with Eq. \eqref{eq:temperature} for a laser to electron coupling efficiency $\eta = 0.51$ (also consistent with our measured laser absorption). For the same target profile (with a relativistically corrected peak density), $\Gamma$ and $\Theta$ are fixed and so is $M_{cr}$, thus the proton energy will depend mainly on the electron temperature. This is confirmed by the measured proton energy scaling with $a_0$, which is in good agreement with Eq. \eqref{eq:scaling}, for an acceleration time of $t_{acc} = 5500 \omega_0^{-1}$ (consistent with the average acceleration time in our simulations (Fig. \ref{fig:scaling} b)). At low intensities the acceleration is dominated by shock reflection (first and second terms of Eq. \eqref{eq:scaling}), but at higher intensities the contribution from the ion expansion (third term of Eq. \eqref{eq:scaling}) also becomes important, leading to a transition from a scaling with $a_0^{3/2}$ to $a_0^2$. This favorable scaling allows for the generation of high quality $\sim 200$ MeV proton beams, required for medical applications \cite{bib:linz}, with a 100 TW class laser system ($a_0 = 10$).

In conclusion, we have presented a scheme for the generation of monoenergetic ion beams. Ions are accelerated by an electrostatic shock driven in an exponentially decaying plasma with a peak density close to critical density. The interaction of an intense laser pulse with such plasmas results in density steepening and strong electron heating which facilitates the formation and stable propagation of a moderate Mach number collisionless shockwave. The narrow spectrum of ions reflected from such a shock is preserved in this tailored plasma profile since the sheath field is constant and small. The high-quality and favorable scaling of the process with laser intensity pave the way for the generation of the ion beams required for medical applications with readily available laser systems.

\begin{acknowledgements}
The authors would like to thank T. Grismayer for discussions. Work supported by the European Research Council (ERC-2010-AdG Grant 267841) and FCT (Portugal) grants PTDC/FIS/111720/2009 and SFRH/BD/38952/2007, and by DOE grant DE-FG02-92-ER40727 and NSF grant PHY-0936266 at UCLA. Simulations were performed at the Jugene supercomputer (Germany) under a PRACE grant, the IST cluster (Lisbon, Portugal), and the Hoffman cluster (UCLA).
\end{acknowledgements}


\newpage

\end{document}